\documentclass[prl,aps,twocolumn]{revtex4}
\usepackage{epsfig,latexsym,amssymb,amsmath,amsbsy,graphics,graphicx}

\begin{document}

\title{Quantum Liquid Crystal Phases in Fermionic Superfluids with Pairing between
Fermion Species of Unequal Densities}

\author{Kun Yang}

\affiliation{National High Magnetic Field Laboratory and Department
of Physics, Florida State University, Tallahassee, Florida 32306,
USA}

\date{submitted to Phys. Rev. Lett. on July 25, 2005)}

\begin{abstract}

Superfluidity in fermionic systems originates from pairing of fermions,
and Bose condensation of these so-called Cooper pairs. The Cooper pairs
are usually made of fermions of different species;
thus the most favorable situation for pairing and superfluidity is
when the two species of fermions that form pairs have the same
density. This paper studies the possible superfluid states when the
two pairing species have {\em different} densities, and show that
the resultant states have remarkable similarities to the phases of
liquid crystals. This enables us to provide a unified description of
the possible pairing phases, and understand the phase transitions
among them.

\end{abstract}

\maketitle

In the Bardeen-Cooper-Schrieffer (BCS) theory for
fermionic superfluidity, fermions form
Cooper
pairs that condense into the zero momentum state.
A Cooper pair is often made of fermions
of different species; in superconductors they are electrons of
opposite spins. Thus the most favorable situation for pairing is when
the two species of fermions have the same density, so that there is no unpaired
fermion in the ground state.
It has been a long-standing fundamental
question as to what kind of pairing states fermions can
form when the two fermion species have different densities or chemical
potentials.  An early suggestion was due to Fulde and Ferrell\cite{ff}, and
Larkin and Ovchinnikov\cite{lo},
who argued that the Cooper pairs may condense into
a finite momentum state; this Fulde-Ferrell-Larkin-Ovchinnikov
(FFLO) state breaks translation and rotation symmetries.
More recently other suggestions have been
put forward, including deformed Fermi surface pairing\cite{ms,ms1,ms2} and
breached pairing\cite{liu,glw,forbes} states,
each with their distinct symmetry properties.
Here we show that all these states are different phases of
a quantum liquid crystal. This allows us to provide a unified description of
all these phases, propose a global phase diagram,
and understand the nature of the transitions between neighboring phases.

The issue of pairing between unbalanced fermion species originally arose in
superconductors subject to either an external magnetic field or internal
exchange field, which gives rise to Zeeman splitting between up- and down-spin
electrons that form Cooper pairs. More recently the same issue has been under
close scrutiny in the context of pairing and superfluidity in nuclear matter\cite{ms1},
neutron stars\cite{ms1}, and high density quark matter\cite{ms2,glw,review},
where the unbalance is due to
more intrinsic effects like difference in mass of the fermions that form pairs.
Perhaps the most promising place where some of these novel phases
can be directly observed are trapped cold atom systems\cite{machida,sedrakian,yang05}, where one
has the best control of the properties of the constituents and
strength of pairing interactions. Thus this is a fundamental issue that is of
importance to {\em all} branches of physics. To better elucidate the properties
of and relations between the various proposed phases, in this work we will
use concepts developed in studies of liquid crystals, which is traditionally
viewed as a branch of {\em classical} physics. The analogy to
liquid crystals allows us to determine the
global phase diagram of the system, and gain insight into the nature of the
transitions between neighboring phases.

We start our discussion with the FFLO state, which has the longest history of
studies, and very strong experimental evidence for its existence has been found
recently in a heavy fermion superconductor, CeCoIn$_5$\cite{radovan,bianchi}.
Following the superconductivity terminology, throughout this paper
we will use ``spin" indices $\sigma=\uparrow, \downarrow$ to label the
two different species of fermions that form Cooper pairs, ``Zeeman splitting"
$\Delta\mu=\mu_\uparrow-\mu_\downarrow$
to represent the chemical potential difference between the two species, and
``magnetization" $m$ to represent their density difference.
When $\Delta\mu\ne 0$, up- and down-spin
electrons form Fermi seas with different Fermi momenta $p_{F\uparrow}$ and
$p_{F\downarrow}$
in the normal state; it was thus suggested\cite{ff,lo}
that when pairing interaction is
turned on, the electrons with opposite spins on their respective Fermi surfaces
would pair up to form a Cooper
pair with a net momentum $p\approx p_{F\uparrow}-p_{F\downarrow}$. This
results in a pairing order parameter $\Delta({\bf r})$ that is oscillatory
in real space, with period $2\pi\hbar/p$. In general the structure of
$\Delta({\bf r})$ is characterized not just by a single momentum $p$, but also
by its higher harmonic components. More detailed mean-field study\cite{rainer}
suggested the following real space picture for FFLO state: it is a state with
a finite density of uniformly spaced domain walls; across each domain wall
the order parameter $\Delta$ (which is real in the mean-field theory)
changes sign, and the excess magnetization due to
spin unbalance are localized along the domain walls, where $\Delta$
(which is also the gap for unpaired fermions) vanishes; see Fig. 1a.
Thus the total magnetization
is proportional to the domain wall density. This picture was made more
precise by an exact solution in one-dimension (1D)
based on bosonized description of
spin-gapped Luttinger liquids\cite{yang01},
where the domain walls are solitons of the sine-Gorden model that describes
the spin sector; each soliton carries one half-spin. While quantum and
thermal fluctuations do not allow true long-range order in 1D, such order
can be stabilized by weak interchain couplings\cite{yang01}.
Coming back to isotropic high D cases, it is clear that the presence and ordering
of these domain walls break rotation symmetry, and  translation symmetry in
the direction perpendicular to the walls, although translation symmetry
along the wall remains intact. Thus the symmetry properties of the FFLO state
is identical to that of the smectic phase of liquid crystals
(smectic-A phase to be more precise)\cite{degennes}.

\begin{figure}
\includegraphics[scale=0.4]{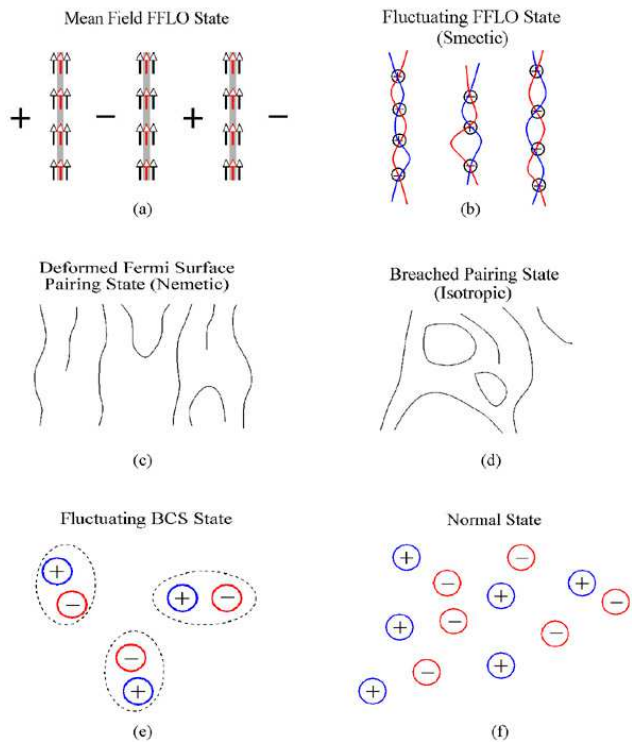}
\caption{(color online). Illustrations of various possible pairing
states. {\bf a}, Domain walls in the mean-field description of the
Fulde-Ferrell-Larkin-Ovchinnikov (FFLO) state, where the pairing
order parameter (or gap) changes sign and the magnetization is
located. {\bf b}, Domain walls of the FFLO state are replaced by
chains of alternating vortices and antivortices once fluctuations
are taken into account. The red and blue lines are where the real
and imaginary parts of the order parameter vanish respectively, and
their intersections are the vortices. The ordered chains break
translation and rotation symmetry in a manner identical to the
smectic phase of a liquid crystal. {\bf c}, The nematic phase of the
vortex/antivortex chains (or their segments, represented by black
lines), which correspond to the deformed Fermi surface pairing
(DFSP) state. {\bf d}, The isotropic phase of the vortex/antivortex
chains (or their segments), which correspond to the breached pairing
(BP) state. {\bf e}, The BCS state in which vortices and
antivortices (generated by thermal or quantum fluctuations) form
closely bound pairs. {\bf f}, The normal state in which the vortices
and antivortices are unbound. }
\end{figure}

The domain wall picture described above needs to be modified when
fluctuations above mean-field theory are taken into account. In
mean-field theory the pairing order parameter $\Delta$ is real; it
thus vanishes along domain walls which are lines in 2D and sheets in
3D. In the presence of fluctuations however $\Delta$ is complex,
thus it only vanishes where the real and imaginary parts of $\Delta$
are {\em simultaneously} zero (see Fig. 1b); these points in 2D and
lines in 3D are nothing but the familiar vortices. Thus in the
presence of fluctuations the mean-field domain walls are unstable
and replaced by chains that are made of alternating vortices and
antivortices (see Fig. 1b), and the FFLO state is a state in which
these chains are generated by the Zeeman splitting, and they line-up
to break the translation symmetry perpendicular to the chain
direction\cite{note}.

The appearance of these vortex-antivortex chains when the fermionic
superfluids are subject to a sufficiently large Zeeman splitting can
also be understood from the following consideration. Zeeman
splitting tends to generate spin-polarized quasiparticles; in a
uniform superfluid state this can happen only when the splitting is
large enough to overcome the quasiparticle gap. On the other hand it
is known\cite{caroli} that the quasiparticle excitations are gapless
inside the vortex core; thus the vortices are easily polarizable,
making their presence energetically favorable when Zeeman splitting
is present. The net vorticity has to remain zero so that equal
numbers of  vortices and antivortices must be generated, and since
vortices repel/attract each other when they have the same/opposite
vorticity, they naturally form an alternating pattern. This
naturally leads to the vortex chain configurations discussed above.

When quantum and/or thermal fluctuations are sufficiently weak,
these infinitely-long vortex/antivortex chains line up and form an
ordered smectic phase that breaks translation symmetry along the
direction perpendicular to the chains and rotation symmetry, as
illustrated in Fig. 1b; this corresponds to the FFLO state in the
presence of fluctuations. In classical liquid crystals it is known
that as one increases thermal fluctuations, the broken symmetries of
the smectic phase are restored in the following
sequence\cite{degennes}: the translation symmetry is restored first
when the smectic melts into a nematic that breaks the rotation
symmetry only, and then the nematic melts into an isotropic liquid
that has no broken symmetry. We thus expect the same sequence of
phases and phase transitions occur in superfluids with unbalanced
fermion pairing, as we increase the strength of either thermal or
quantum fluctuations. More specifically, as fluctuations increase,
these infinitely long vortex/antivortex chains can break into
segments so that dislocations can appear in the lattice formed by
these chains; when such dislocations proliferate and unbind the
translation symmetry is restored, although (segments of) the chains
are still aligned in a preferred direction, so that the rotation
symmetry remains broken (see Fig. 1c for an illustration). This is
the analog of the nematic phase, which has the same symmetry as the
deformed Fermi surface pairing (DFSP) state\cite{ms,ms1,ms2}; we
thus identify the DFSP state as the nematic. The chains (or their
segments) will lose their orientation alignment upon further
increasing fluctuations (see Fig. 1d for an illustration), thus
restoring the rotation symmetry (through proliferation of
disclinations); this isotropic liquid phase is thus identified with
the breached pairing (BP) phase\cite{liu,glw,forbes}. While all
broken spatial symmetries are restored by fluctuations at this
point, these phases are still superfluid phases with a spontaneously
broken U(1) gauge symmetry, as long as the vortices and antivortices
are bound together to form chains or their segments. Further
increasing fluctuations will eventually break the segments apart
into unbound vortices; this is the normal phase in which the U(1)
gauge symmetry is also restored (see Fig. 1f).

It is clear that the fundamental similarity among the three different
superfluid phases discussed
above is the appearance of the vortex/antivortex chains (or their segments)
in the ground state; they are responsible for accommodating the excess
magnetization or unbalance between the fermion species. This is very different
from the BCS phase. While in the BCS phase (or ordinary superfluids due to
simple Bose condensation as described by the XY model) thermal or quantum
fluctuations can still generate vortex excitations, they appear as bound
{\em pairs} each made of a vortex and an antivortex (see Fig. 1e),
instead of forming chains or
segments of chains.

Based on the discussion above, we propose a phase diagram in 3D
(Fig. 2) for pairing between unbalanced fermion species, assuming
the Zeeman splitting $\Delta\mu$ is strong enough to destabilize the
BCS state, yet not strong enough to eliminate pairing and
superfluidity at $T=0$. For pairing fermions with the same mass,
this requires $\Delta\mu \sim \Delta_0$, where $\Delta_0$ is the
pairing gap at $T=0$ and $\Delta\mu=0$; it parametrizes how much
energy it costs to break a Cooper pair. We use $\Delta_0$ as our
unit of energy (although it depends on the pairing strength and
other parameters itself). As usual thermal fluctuation is controlled
by temperature. Quantum fluctuation is controlled by the strength of
pairing interaction; it can be measured by the dimensionless ratio
$\Delta_0/E_F$, where $E_F=p_F^2/2M$ is the Fermi energy in the
absence of $\Delta\mu$ and pairing interaction, and $M$ is the
fermion mass. In trapped cold fermionic atom systems the pairing
strength can be controlled by tuning the distance to a Feshbach
resonance by a magnetic field, where one can explore a wide
parameter range from the weak coupling (or BCS) to strong coupling
(which corresponds to the Bose-Einstein condensation or BEC of
diatom molecules) regimes of superfluidity\cite{feshbach}. As we
increase thermal (increasing $T$) and/or quantum (increasing
$\Delta_0/E_F$) fluctuations, the smectic phase (that represents the
FFLO state) first melts into a nematic (representing the DFSP
state), and then yields to the isotropic (representing the BP
state). The experimental detection of these phases should be fairly
straightforward in these systems\cite{sedrakian,yang05}. The
nematic-isotropic transition is first order due to a cubic term in
the appropriate Landau theory\cite{degennes}. The
smectic(-A)-nematic transition is (typically) a continuous
transition, although the experimentally observed critical behavior
is not yet completely understood\cite{degennes}. Direct transition
between these superfluid states and the normal state are also
possible. The transition between the BP and normal phases is in the
same universality class as the usual superfluid (or XY) transition,
which is usually continuous. On the other hand the FFLO/DFSP to
normal transitions should be first order even if the mean-field
theory suggests a continuous transition; this is because both
spatial and gauge symmetries are broken simultaneously here. Such
fluctuation-driven first-order transition has been explicitly
demonstrated for the FFLO case recently\cite{denis04}. The situation
is quite different in 2D; there the smectic phase is unstable for
any finite temperature, while the nematic-isotropic transition is
expected to be of the Kosterlitz-Thouless type\cite{tn}. On the
other hand the quantum (or $T=0$) phase transitions among these
phases are poorly understood at this point, and will be the subject
of future work.

\begin{figure}
\includegraphics[scale=0.35, angle=90]{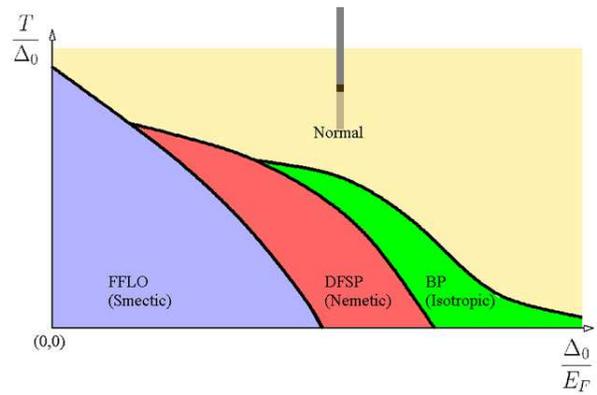}
\caption{(color online). Schematic phase diagram for pairing between
unbalanced fermion species in 3D. The FFLO state, which is a
smectic, is stable at low temperature and weak coupling. As thermal
fluctuation (controlled by $T$) and quantum fluctuation (controlled
by pairing strength and parametrized by the ratio between pairing
gap and Fermi energy, $\Delta_0/E_F$) increase, it first melts into
a nematic (or the deformed Fermi sea pairing state), which then
melts into an isotropic liquid crystal (or breached pairing state).
Direct transitions between these superfluid states and the normal
state are also possible at high temperature. }
\end{figure}

In the following we present a concrete case of the breached pairing (BP) state,
in the strong-coupling (or BEC) regime. In this case unpaired, spin-polarized
fermions co-exist with closely-bound fermion pairs (or molecules) that Bose
condense; the appropriate fermionic many-body wave function takes the form:
\begin{equation}
|\Psi\rangle=\prod_{|{\bf k}| \le k_F'}c^\dagger_{{\bf k},\uparrow}
\prod_{|{\bf k}| > k_F'}(u_{\bf k}
+v_{\bf k}c^\dagger_{{\bf k},\uparrow}c^\dagger_{-{\bf k},\downarrow})|0\rangle,
\label{bp}
\end{equation}
where $|0\rangle$ is the vacuum state,
$k_F'=[6\pi^2(n_\uparrow-n_\downarrow)]^{1/3}$ is the Fermi
wave-vector of the {\em excess} (or unpaired) fermions, and $u_{\bf
k}$ and $v_{\bf k}$ (which satisfy $u_{\bf k}^2+v_{\bf k}^2=1$) are
coherence factors; they are determined by identifying $\Phi_{\bf
k}=v_{\bf k}/u_{\bf k}$ as the (momentum space) molecular wave
function\cite{leggett}, which in turn can be obtained by solving the
two-body Schrodinger's equation with the constraint that $\Phi_{\bf
k}=0$ for $|{\bf k}| \le k_F'$:
\begin{equation}
{\hbar^2k^2\over M}\Phi_{\bf k} + \sum_{\bf k'}V_{{\bf k}{\bf
k}'}\Phi_{\bf k'}=-2|\Delta|\Phi_{\bf k},
\end{equation}
where $V_{{\bf k}{\bf k}'}$ is the pairing potential in momentum
space. This only leads to a very minor modification of $\Phi_{\bf
k}$ from the genuine two-body problem in the BEC limit, because
components with $|{\bf k}| \le k_F'$ only carry a very small weight
for the closely bound state with size much smaller than $1/k_F'$;
thus the pairing is only ``breached slightly" by the presence of the
spin-polarized, unpaired fermions. Clearly this is the energetically
favored state at strong coupling (compared to the FFLO and DFSP
states), as it minimizes the kinetic energy of the unpaired fermions
while preserve the pairing correlation for the remaining fermions
that do form pairs. The wave function (\ref{bp}) takes a form
similar to the one proposed by Liu and Wilczek\cite{liu}; the main
difference is that here the unpaired fermions occupy states near the
origin in momentum space (instead of near $k_F$ of the
non-interacting Fermi gas\cite{liu}) and form a circular Fermi sea
of itself; the origin of the difference is we are considering the
strong coupling case while Liu and Wilczek were considering the
weak-coupling limit. But they share the same symmetry property (both
isotropic), and are therefore different forms of the BP phase.

We close by noting that quantum liquid crystal phases have also been
proposed to describe other strongly interacting fermionic systems,
including cuprate superconductors\cite{kfe} and quantum Hall
liquids\cite{fk,radzihovsky}.

 We thank Nick Bonesteel,
 Eugene Demler, Akakii Melikidze, Jorge Piekarewicz, and Leo Radzihovsky
for useful discussions, and Eddy Yusuf for technical assistance.
This work was supported by by National Science Foundation grant No.
DMR-0225698.

{\em Note Added} -- After the present paper has been submitted for
publication, the author became aware of three interesting new
preprints\cite{pao,son,sheehy} that address closely related issues
from somewhat different angles.



\begin{thebibliography}{1}
\bibitem{note} Under certain circumstances the vortices may also order along the
chain direction and break translation symmetry in that direction
too; we do not consider the possibility of such (possibly
anisotropic) crystal phases here.
\end{thebibliography}
\end{document}